\documentclass[pra,aps,twocolumn,showpacs]{revtex4}
\usepackage{graphicx}
\usepackage{dcolumn}
\newcommand{\nn}{\nonumber}

\begin{document}
\def\be{\begin{equation}}
\def\ee{\end{equation}}
\def\bearr{\begin{eqnarray}}
\def\eearr{\end{eqnarray}}
\def\la{\langle}
\def\ra{\rangle}
\def\l{\left}
\def\r{\right}

\title{Bragg spectroscopy of a cigar shaped Bose condensate in optical lattices}

\author{Tarun Kanti Ghosh$^{1,2}$ and Kazushige Machida$^{1}$}
\affiliation
{$^{1}$Department of Physics, Okayama University, Okayama 700-8530, Japan\\
$^{2}$Institute for Theoretical Physics, Heinrich-Heine Duesseldorf University, 
40225, Duesseldorf, Germany }

\date{\today}

\begin{abstract}
We study properties of excited states of an array of weakly 
coupled quasi-two-dimensional Bose condensates by using the 
hydrodynamic theory. We calculate multibranch Bogoliubov-Bloch 
spectrum and its corresponding eigenfunctions. The spectrum 
of the axial excited states and its eigenfunctions strongly 
depends on the coupling among various discrete radial modes 
within a given symmetry. This mode coupling is due to the
presence of radial trapping potential.
The multibranch nature of the Bogoliubov-Bloch spectrum and 
its dependence on the mode-coupling can be realized by analyzing 
dynamic structure factor and momentum transferred to the system 
in Bragg spectroscopy experiments.
We also study dynamic structure factor and momentum transferred to 
the condensate due to the Bragg spectroscopy experiment. 

\end{abstract}

\maketitle

\section{Introduction} 
The experimental realization of optical lattices \cite{op} is
stimulating new perspectives in the study of strongly correlated 
cold atoms \cite{rmp}. 
Bose-Einstein condensates (BEC) in optical lattices are promising
realistic systems to study the superfluid properties of Bose gases
\cite{op4,op5,ichioka}. 
When height of the lattice potential is low, the BEC have coherency over all
lattice sites. However, when height of the lattice potential increases
so that nearest neighbor site hopping becomes difficult for atoms and
then coherence of the BEC is destroyed. The atomic condensate placed
in a three-dimensional (3D) optical lattice can be best described by the 
Bose-Hubbard model \cite{bh1}.  
The Bose-Hubbard model has been realized and 
quantum phase transition from weakly interacting Bose superfluid 
to a strongly interacting Mott insulator state was indeed observed 
experimentally \cite{bh2,bh3}. Similar quantum phase transition was 
also observed when a 3D BEC was placed in a one-dimensional (1D) optical
lattice \cite{bh4}. The insulator phase in the 1D lattice is called a squeezed state.
Kramer {\em et al.} \cite{kramer} have found the mass renormalization 
in presence of the optical potential.
The renormalization decreases the value of the 
axial excitation frequencies which have been verified experimentally 
\cite{kramer1}. 
There are several theoretical calculations for dynamic
structure factor of quasi-one-dimensional (1D) Bose gas placed 
in 1D optical lattices \cite{dsf1,dsf2,dsf3}.
Recently, Zobay and Rosenkranz \cite{zobay} have studied effect of weak 1D 
periodic potential on the thermodynamic properties (critical temperature,
critical number and critical chemical potential) of a homogeneous interacting
Bose gases by using mean-field and renormalization group method.
Sound velocity in an interacting Bose system in presence of periodic optical
lattices has been studied by various authors \cite{molmer,java,pethick,taylor}.

A stack of weakly coupled quasi-two-dimensional (quasi-2D) condensates 
can be created by applying a relatively strong 1D 
optical lattices to an ordinary 3D cigar shaped condensate. 
The finite transverse size of the condensates produces a discreteness 
of radial spectrum. The short wavelength axial phonons with different 
number of discrete radial modes of a cigar shaped condensates 
placed in optical lattices give rise to multi-branch Bogoliubov-Bloch 
spectrum (MBBS) \cite{tkg}. This is similar to multibranch Bogoliubov 
spectrum (MBS) of a cigar shaped BEC \cite{mbs1,fedichev}. 

Bragg spectroscopy of a trapped BEC has proven to 
be an important tool for probing many bulk properties such as 
dynamic structure factor \cite{phonon}, verification of multibranch 
Bogoliubov excitation spectrum \cite{mbs2}, 
momentum distribution and correlation functions of a phase fluctuating 
quasi-1D Bose gases \cite{ric,gerbier} and even the velocity field of a 
vortex lattice within a condensate \cite{raman}.

Recently, we have studied the MBBS by including couplings among all
low energy modes in the same angular momentum sector by using 
hydrodynamic theory \cite{tkg}. This mode coupling is due to the presence 
of inhomogeneous radial density. We should mention that the MBBS for 
phonon and breathing modes have been studied by including couplings between
phonon and breathing modes only by using 
variational method \cite{stoof1}. The mode couplings among all low energy
modes must be considered in order to calculate correct eigen spectrum and the
corresponding eigen functions \cite{tkg}. 
The MBBS softens as we increase strength
of the optical potential. It indicates a transition from the superfluid state
to the Mott insulating state along the optical lattice but the superfluid
state remains in each quasi-2D condensates.

To understand effect of the mode coupling on the spectrum and 
the softening of these modes as we increase the laser intensity, we need to measure 
those modes. 
There is no theoretical study on how to measure these modes in the superfluid regime.
The MBBS can be measured by using the Bragg scattering experiments 
via dynamic structure factor measurements. 
In this work, we start from the beginning systematically by presenting results of 
the MBBS and the corresponding wave functions of the density fluctuations.
Then we calculate the dynamic structure factor and momentum transferred
to the system which will give us information about the spectrum.

This paper is organized as follows. In Sec. II, we consider a stack
of weakly coupled quasi-two-dimensional Bose condensates. Using the discretized
hydrodynamic theory, we calculate the multibranch Bogoliubov-Bloch spectrum and
its corresponding eigenfunctions by including the mode coupling within a given 
symmetry. In Sec. III, we study the dynamic structure factor.
In Sec. IV, we analyze the momentum transferred to the condensates by Bragg pulses. 
We also calculate time duration needed for the Bragg potential to observe 
the multibranch nature of the spectrum.
We give a brief summary and conclusions in Sec. IV.

\section{MBBS of a stack of Bose condensates}
We consider that bosonic atoms, at $T=0$, are trapped by a
cigar shaped harmonic potential and a stationary 1D optical potential 
modulated along the $z$ axis. The Gross-Pitaevskii energy functional 
can be written as
\bearr
E_0 & = & \int dV \psi^{\dag}(r,z)[- \frac{\hbar^2}{2M} {\bf \nabla}^2 + 
V_{\rm ho}(r,z) 
\nonumber \\ & + &  
\frac{g}{2} |\psi(r,z)|^2 + V_{\rm op}(z)] \psi(r,z).
\eearr
Here, $ V_{\rm ho}(r,z) = (M/2)(\omega_r^2 r^2 + \omega_z^2 z^2) $ is the
harmonic trap potential and
$ V_{\rm op}(z) = s E_r \sin^2(q_0z) $ is the optical potential modulated
along the $z$ axis where $ E_r = \hbar^2 q_0^2/2M $
is the recoil energy, $ s$ is the dimensionless parameter determining the
laser intensity and $ q_0 $ is the wave vector of the laser beam. Also,
$ g = 4 \pi a \hbar^2/M $ is the strength of the two-body interaction energy, 
where $a$ is the two-body scattering length.
We also assumed that $ \omega_r >> \omega_z $ so that it makes
a long cigar shaped trap. The minima of the optical potential are located
at the points $ z_j = j \pi/q_0 = j d $, where $d = \pi/q_0 $ is the
lattice size along the $z$ axis. Around these minima, the optical 
potential can be approximated as 
$ V_{\rm op}(z) \sim (M/2)\omega_s^2 (z-z_j)^2 $, 
where the layer trap frequency is $ \omega_s = \sqrt{s} \hbar q_0^2/M $. 
In the usual experiments, the well trap frequency is larger than 
the axial harmonic frequency, $  \omega_s >> \omega_z $. 
For a typical experiment \cite{kramer1}, the system parameters are 
$ \omega_r \simeq 2\pi \times 90 $ Hz, $ \omega_z \simeq 2\pi \times 9 $ Hz, and
$ \lambda = q_0/2\pi = 820 $ nm. 
When $  \omega_s >> \omega_z $ and the number of quasi-2D layers is quite large,
we can safely ignore the harmonic potential along the $z$ axis.
Moreover, we also consider the atom number in each quasi-2D condensates to be
constant and equal at each layer. The number of atoms in each layer for a typical
experiment is $ N \sim 10^4 $.

The strong laser intensity will give rise to a stack
of several quasi-two-dimensional condensates. Because of quantum 
tunneling, the overlap between the wave functions of two
consecutive layers can be sufficient to ensure full coherence.
If the tunneling is too small, the strong phase fluctuations 
will destroy axial phase coherence among the layers and lead to a 
new strongly correlated quantum state, namely Mott insulator state
along the $z$ axis and superfluid state in each quasi-2D condensate.
One can give a rough estimate of the threshold laser intensity
for the exotic quantum phase transition to be quite large \cite{stoof}.  

In the presence of coherence among the layers it is natural 
to make an ansatz for the condensate wave function as 
\be \label{dwf} 
\psi(x,y,z) = \sum_j \psi_j(x,y) f(z-z_j).
\ee
Here, $ \psi_j(x,y) $ is the wave function of the quasi-two-dimensional
condensate at a site $j$ and $ f(z-z_j) $ is a 
localized function at the $j$-th site.
The localized function can be written as
\be \label{ansatz}
f(z-z_j) = (\frac{M \omega_s}{\pi \hbar})^{1/4} 
e^{-(M \omega_s/2 \hbar) (z-z_j)^2}.
\ee
Substituting the above ansatz into the energy functional and
considering only the nearest-neighbor interactions, one can 
get the following energy functional:
\bearr \label{main}
E_0 & \simeq & \sum_j \int dx dy [-\frac{\hbar^2}{2M} 
\psi_j^{\dag} {\bf \nabla}_r^2 \psi_j
+ V_{\rm ho}(r) |\psi_j|^2]
\nonumber \\ & + &
\frac{ g_{\rm 2D}}{2} \sum_j \int dx dy \psi_j^{\dag} \psi_j^{\dag} \psi_j \psi_j
\nonumber \\ & - &
J \sum_{j,\delta = \pm 1} \int dx dy [\psi_{j+\delta}^{\dag} \psi_j +  \psi_j^{\dag} 
\psi_{j+\delta}],
\eearr
where $ J $ is the strength of the Josephson coupling 
between adjacent layers which is 
given as
\bearr
J & = & -  \int dz   f(z) [- \frac{\hbar^2}{2M}   \nabla_z^2 +
V_{\rm op}(z) ] f(z + d) \nonumber \\
& \sim & \hbar \omega_r (\frac{ \pi a_r}{\sqrt{2} \lambda})^2 (\pi^2 - 4)s 
e^{-(\pi^2 \sqrt{s}/4)},
\eearr
where $ a_r = \sqrt{\hbar/M \omega_r} $. 
Also, the strength of the effective on-site interaction energy is
$  g_{\rm 2D} = g \int dz |f_0(z)|^4 = 4 \sqrt{\pi/2}
(\hbar^2/M)(a/a_s)$,
where $ a_s = \sqrt{\hbar/M \omega_s} $.
Equation (\ref{main}) shows that each layer $j$ is coupled with the
nearest-neighbor layers $ j\pm 1$ through the tunneling energy $J$.
The axial dimension appears through the Josephson coupling between 
two adjacent layers. 
The Hamiltonian corresponding to the above energy functional is
similar to an effective 1D Bose-Hubbard Hamiltonian in which each lattice
site is replaced by a layer with inhomogeneous radial density.

Using phase-density representation of the bosonic field operator as
$ \psi_j = \sqrt{n_j}e^{i\theta_j} $ and
following Ref. \cite{tkg}, equations of motion for the density and phase 
fluctuations can be written as
\bearr \label{dfluc}
\delta \dot n_j & =  &- \frac{\hbar}{M} {\bf \nabla}_r \cdot [n_{0}(r) {\bf \nabla}_r \delta 
\theta_j] 
\nonumber \\
& + &  \frac{2J}{ \hbar} n_{0}(r) [2 \delta \theta_j -  \delta \theta_{j-1} - \delta 
\theta_{j+1}]
\eearr
and
\be \label{vfluc}
\hbar \delta \dot \theta_j = -  g_{\rm 2D} \delta n_j - \frac{J}{2n_0(r)}
[2\delta n_j - \delta n_{j-1} - \delta n_{j+1}].
\ee
Here, $ ``{}^{\cdot}" $ represents the time derivative. In equilibrium, the condensate
density at each layer within the Thomas-Fermi (TF) approximation is $ n_{0}(r) \simeq 
[\mu_0 - V_{\rm ho}(r)]/ g_{\rm 2D}$,
where we have neglected the effect of the tunneling energy $J$ since it is very
small in the deep optical lattice regime.
Also, $ \mu_0 = \hbar \omega_r \sqrt{\sqrt{8/\pi}(Na/a_s)} $ is the
chemical potential at each layer, where $N$ is the number of atoms
at each layer.
Note that the second term ($ \epsilon (r) = J/2n_0(r) $) of the right hand side of Eq. 
(\ref{vfluc}) is proportional to the small parameter $J$ and inversely proportional
to the large parameter $ n_0(r =0) = \mu_0/ g_{\rm 2D} $. In our previous
publication \cite{tkg}, we have neglected this small term. Within the TF
regime, this term is really negligible. 
However, to present more accurate result we keep this term by approximating safely as 
$ J/2n_0(r) \sim J/2n_0(r=0) = J g_{\rm 2D}/\mu_0 $. One may raise a objection that
this approximation is not valid because the ratio $ \epsilon (r) $ is large as
we go away from the trap center and it diverges at the boundary of the system. 
This divergence at the boundary is due to an artifact of the TF approximation to the 
density profile.  
The main idea of the linear response theory is to satisfy the following relation: 
$ \delta n (r) /n_0(r) << 1 $. In our case, the term $ J \delta n/2n_0(r) $ is
always small within the linear response theory.  

The density and phase fluctuations can be written in a canonical form as
\be \label{dencan}
\delta \hat n_{j}(r,t) = \sum_{\alpha,k} [A_{\alpha,k} \psi_{\alpha,k}({r})
\hat b_{\alpha,k}
e^{i(jkd-\omega_{\alpha}(k)t)} + h. c.],
\ee
and
\be \label{phasecan}
\delta \hat \theta_{j}(r,t) = \sum_{\alpha,k} [B_{\alpha,k} \psi_{\alpha,k}({r}) 
\hat b_{\alpha,k}e^{i(jkd- \omega_{\alpha}(k)t)} + h. c.].
\ee
Here, $\hat b $ is a destruction operator of a quasiparticle and 
$\alpha$ is a set of two quantum numbers: radial quantum number,
$n_r$ and the angular quantum number, $m$. Also, $k$ is Bloch wave vector of the
excitations. The Bloch wave vector $p$ which is associated with the
velocity of the condensate in the optical lattice is set to zero.
The density and phase fluctuations satisfy the following equal-time commutator
relation: $  [\delta \hat n(r), \delta \hat \theta(r^{\prime})]
= i \delta (r - r^{\prime})$.
One can easily show that
\be
A_{\alpha,k} = i \sqrt{\frac{\hbar \omega_{\alpha}(k)}{2 g_{\rm 2D} B_0}},
\hspace{0.3cm}
B_{\alpha,k} = \sqrt{\frac{ g_{\rm 2D} B_0}{2\hbar \omega_{\alpha}(k)}},
\ee
where $ B_0 = 1+(2J/\mu_0) \sin^2(kd/2) $ is a dimensionless parameter.
The parameter $ B_0 \simeq 1$ within the TF regime as well as in the tight 
binding regime.
We are assuming that $ \psi_{\alpha,k}(r) $ satisfies the orthonormal
conditions:
$ \int d^2 r \psi_{\alpha,k}^*(r) \psi_{\alpha^{\prime},k}(r) =
\delta_{\alpha \alpha^{\prime}} $
and $ \sum_{\alpha} \psi_{\alpha,k}^*(r) \psi_{\alpha,k}(r^{\prime}) =
\delta (r -  r^{\prime}) $.

Using Eqs. (\ref{dfluc}), (\ref{vfluc}), (\ref{dencan}), and (\ref{phasecan}), we get
\bearr \label{density1}
-\omega_{\alpha}^2(k) \psi_{\alpha,k} & = & 
B_0 \frac{ g_{\rm 2D}}{M} {\bf \nabla}_r \cdot [n_{0}(r) 
{\bf \nabla}_r \psi_{\alpha,k}] \nonumber \\
& - & B_0 \frac{8 J g_{\rm 2D}}{\hbar^2} n_{0}(r) \sin^2(kd/2)  \psi_{\alpha,k}.
\eearr
  
For $ k =0 $, the solutions are known exactly and analytically
\cite{graham}.
The energy spectrum and the normalized eigen functions, respectively, are given as
$ \omega_{\alpha}^2 = \omega_{\alpha}^2(k=0) = \omega_r^2[|m| + 2 n_r(n_r +|m| +1)] $ 
and
$$ \nonumber 
\psi_{\alpha}(r,\phi) =\psi_{\alpha,k=0}(r,\phi) \sim \nn \tilde r^{|m|} 
P_{n_r}^{(|m|,0)}(1-2 \tilde r^2) e^{i m \phi} \nonumber.
$$
Here, $ P_n^{(a,b)}(x) $ is the Jacobi polynomial of order $n$ and $ \phi $ is the 
polar angle. The radius of each condensate layer is $ R_0 = 2 \mu_0/M \omega_r^2 $ and
$ \tilde r = r/R_0 $ is the dimensionless variable. 

The solution of Eq. (\ref{density1}) can be obtained for arbitrary value of $k$ by
numerical diagonalization. 
For $ k \neq 0 $, we can expand the density fluctuations as
\be
\psi_{\alpha,k} (r,\phi) = \sum_{\alpha} c_{\alpha} \psi_{\alpha}(r,\phi).
\ee 
Substituting the above expansion into Eq.(\ref{density1}), we obtain,
\bearr \label{density2}
0 & = & [\tilde \omega_{\alpha}^2(k) - B_0[|m| + 2 n_r(n_r +|m| +1)]] c_{\alpha} 
\nonumber \\ & - &
B_0 B_1\sin^2(kd/2) \sum_{\alpha^{\prime}} M_{\alpha \alpha^{\prime}} 
c_{\alpha^{\prime}},
\eearr 
where $ \tilde \omega_{\alpha} = \omega_{\alpha}/\omega_r $ and the dimensionless 
parameter $B_1$ is defined as
$B_1 = 8 \tilde J \tilde \mu_0 $. Here, $ \tilde J = J /\hbar \omega_r $
and $ \tilde \mu_0 = \mu_0 /\hbar \omega_r $. 
The matrix element $ M_{\alpha \alpha^{\prime}} $ is given by
\be\label{matrix}
M_{\alpha \alpha^{\prime}}  =  \int d^2 \tilde r 
\psi_{\alpha}^*(r,\phi) (1 - \tilde r^2) \psi_{\alpha}(r,\phi).
\ee
The above eigenvalue problem is block diagonal with no overlap
between the subspaces of different angular momentum, so
that the solutions to Eq.(\ref{density2}) can be obtained separately in
each angular momentum subspace. We can obtain all low energy
multibranch Bogoliubov-Bloch spectrum and the corresponding eigenfunctions 
from Eq. (\ref{density2}). Equations (\ref{density2}) and (\ref{matrix}) show 
that the spectrum depends on average over the radial coordinate and the 
coupling among the modes within a given angular momentum symmetry for any 
finite value of $k$.
Particularly, the couplings among all other modes are important for large
values of $kd $ and $ B_1$. 
It is interesting to note that the curvature of a spectrum strongly depends
on the parameter $ B_1 $ and weakly depends on the parameter $B_0$.
The tunneling energy ($J$) decreases as we increase the strength of the
optical potential and therefore the modes become soft. 

When the radial trapping potential is absent, we obtain the usual  
Bogoliubov-Bloch spectrum from Eq. (\ref{density2}) and it is 
given by
\be
\tilde \omega_{\rm homo}^2(k) = 8 \tilde J \tilde \mu_0 \sin^2(kd/2) + 
16 \tilde J^2 \sin^4(kd/2).
\ee
In the long-wavelength limit, the sound velocity is given by
$ c_{\rm homo} = \sqrt{\mu_0/M^*} $, where 
$ M^* = \hbar^2/2 J d^2 $ is the effective mass of 
the atoms in the optical potential. 

Now we consider inhomogeneous radial density.
In the limit of long wavelength, the $ n_r = 0, m = 0 $ mode 
is phonon-like with a sound velocity $c_0 = \sqrt{\mu_0/2M^*} $. 
This sound velocity exactly matches with the result obtained in
Ref. \cite{kramer} and is similar to the result 
obtained without optical potential \cite{mbs1}. 
This sound velocity is smaller by a factor of $ \sqrt{2} $ with
respect to the sound velocity for homogeneous condensate placed
in 1D optical lattices.
This is due to the effect of the average
over the radial variable which can be seen from Eqs. (\ref{density2}) and 
(\ref{matrix}). 

In Fig. 1, we show few low-energy multibranch Bogoliubov-Bloch spectrum 
in the $ m = 0 $ sector as a function of $kd$ by solving the matrix 
Eq. (\ref{density2}) for given values of $ J = 0.1 \hbar \omega_r $ and 
$ \mu_0 = 50 \hbar \omega_r $. 
Hereafter, we will also use these parameters for other figures.
The tunneling energy $J = 0.1 \hbar \omega_r $ corresponds to
the strength of the laser intensity $ s \sim 13 $. Within this parameter regime, 
atoms can tunnel from one layer to the adjacent layer and the whole system 
lies within the superfluid regime \cite{stoof}.
\begin{figure}[ht]
\includegraphics[width=8.0cm]{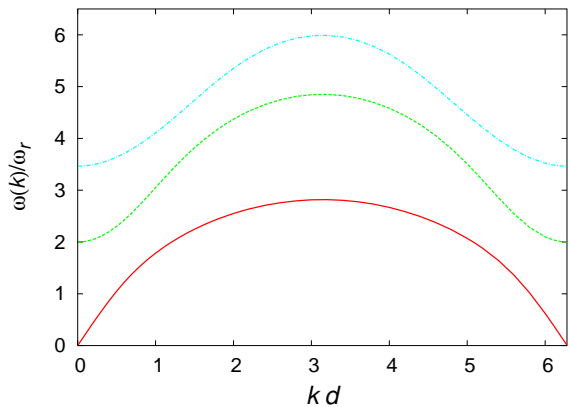}
\caption{(Color online) Plots of the low-energy Bogoliubov-Bloch modes in 
the $m=0$ sector.
Here, $ J =0.1 \hbar \omega_r $ and $ \mu_0 = 50 \hbar \omega_r $.}
\end{figure}
The lowest branch corresponds to the Bogoliubov-Bloch axial 
mode with no radial nodes. This mode has the usual form like 
$ \omega(k) = c_0 k $ at low momenta.
The second branch corresponds to one radial node and starts at 
$ 2 \omega_r $ for $k=0$. The breathing mode has the free-particle 
dispersion relation and it can be written in terms of the
effective mass ($m_b^* $) of this mode as 
$ \omega_2(k) = 2 \omega_r + \hbar k^2/2m_b^* $. 

The eigenfunctions corresponding to the low-energy modes are shown
in Fig. 2. The eigenfunctions also satisfy the symmetry rule:
$ \psi_{n_r,k}(r) = \psi_{n_r,2\pi - k}(r) $. 
\begin{figure}[ht]
\includegraphics[width=8.0cm]{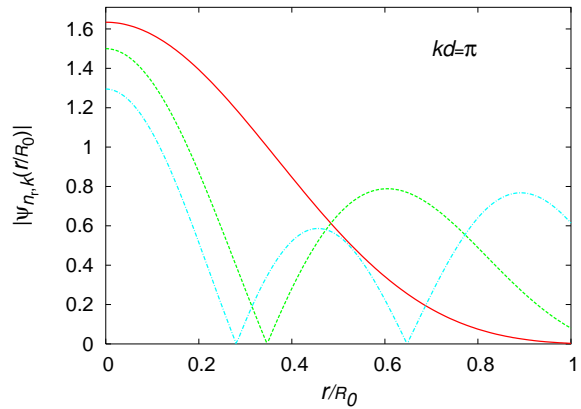}
\caption{(Color online) Plots of the eigenfunctions of the low-energy modes for 
$ J =0.1 \hbar \omega_r $ and $ \mu_0 = 50 \hbar \omega_r $.} 
\end{figure}
\vspace{1cm}
\section{Dynamic Structure Factor}
The dynamic structure factor is the Fourier transformation of density-density
correlation functions. The dynamic structure factor for the layers of
quasi-2D condensates can be written as
\bearr
S(q,\omega) & =  & \sum_{j,j^{\prime}} \int d^2 r d^2 r^{\prime} dt
e^{i qd (j -  j^{\prime})} e^{i\omega t} \nonumber \\
& \times & <\delta \hat n_j^{\dag}(r,t) \delta \hat n_{j^{\prime}} (r^{\prime},0)>.
\eearr
Here, we have used the fact that $ q $ is perpendicular to the
radial plane.
At zero temperature, it can be re-written as
\be
S(q,\omega) =  \sum_{\alpha} S_{\alpha}(q) \delta (\omega - \omega_{\alpha}(q)),
\ee
where weight factor $ S_{\alpha}(q) $ is given by
\be
S_{\alpha}(q) = | A_{\alpha,q}|^2 |\psi_{\alpha}(q)|^2.
\ee
Here, $ \psi_{\alpha}(q) = \int d^2 r \psi_{\alpha,q}({r})$. 
The weight factors are shown in Fig. 3 as a function of $kd$.
The weight factors $ S_{\alpha}(q) $
determine how many modes are excited for a given value of $q$. For example, 
only $ n_r = 1$ and $2$ modes are excited when $ qd = 1.0 $ and other modes
are vanishingly small.
The weight factors also satisfy the lattice symmetry, $ i. e. $, 
$ S_{\alpha}(q) = S_{\alpha}(2\pi - q) $.
\begin{figure}[ht]
\includegraphics[width=8.0cm]{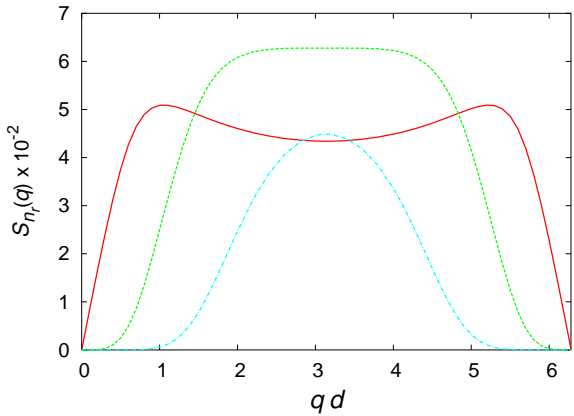}
\caption{(Color online) Plots of the weight factor for $ J =0.1 \hbar \omega_r $ and 
$ \mu_0 = 50 \hbar \omega_r $}. 
\end{figure}
In Fig. 4, we also plot the dynamic structure factor as a function of the Bragg
frequency $\omega $ for two values of the Bragg momenta $q$ with fixed $ J $ and 
$ \mu_0 $.
\begin{figure}[ht]
\includegraphics[width=8.0cm]{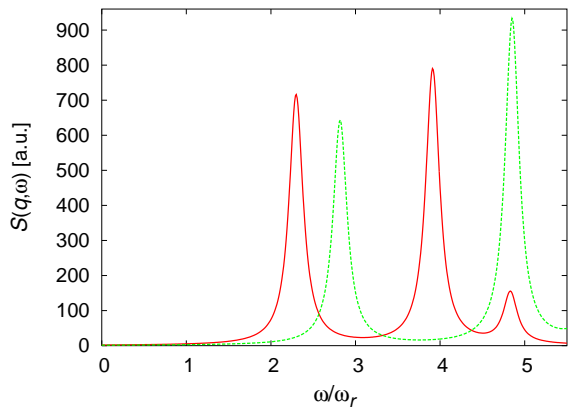}
\caption{(Color online) Plots of the dynamic structure factor for
$kd = \pi/2$ (solid) and $kd = \pi$ (dashed) with fixed
$ J =0.1 \hbar \omega_r $ and $ \mu_0 = 50 \hbar \omega_r $.}
\end{figure}
We find multiple peaks in the dynamic structure factor for a
given value of $q$.
The location of the peaks correspond to the excitation energy for
a given value of $ q $ with fixed $J$ and $ \mu_0 $.
These peaks show the multibranch nature of the Bogoliubov-Bloch modes.

\section{Bragg Spectroscopy}
\begin{figure}[ht]
\includegraphics[width=8.0cm]{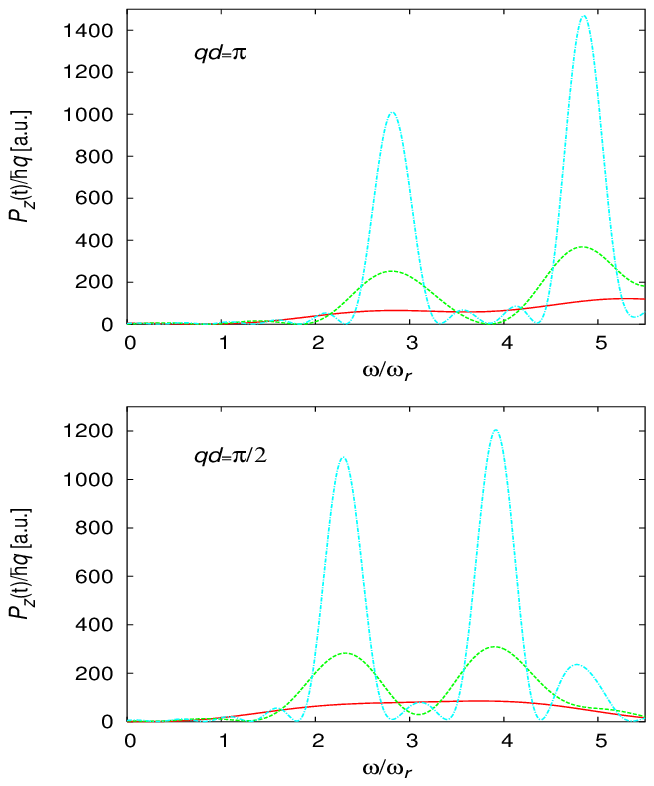}
\caption{(Color online) Plots of the $P_z(t) $ for $ J =0.1 \hbar \omega_r $ and
$ \mu_0 = 50 \hbar \omega_r $ for various times $t = 0.5 T_r $(solid),
$t = 1.0 T_r $ (dashed) and $t = 2.0 T_r $ (dot-dashed).}
\end{figure}
The behavior of these multiple peaks in the dynamic structure factor can
be resolved in a two-photon Bragg spectroscopy, as shown by Steinhauer
{\em et al.} \cite{mbs2}. 
The Bragg scattering experiments can be done by applying an additional 
moving optical potential in the form of $ V_{B}(t) = V_0 \cos(qz-\omega t)$.
Here, $ V_0 $ is the intensity of the Bragg pulses. 
The Bragg potential $ V_{B}(t) $ is independent from the static lattice 
potential $ V_{\rm op} $ and it is also much weaker than the lattice potential. 
We also assume that the 
excitations are confined within the lowest band, therefore, $ \omega $ must 
be less than the energy gap between the first and second bands.
In the two-photon Bragg spectroscopy, the dynamic structure factor can not 
be measured directly. Actually, the observable
in the Bragg scattering experiments is the momentum transferred to the
condensate which is related to the dynamic structure factor and reflects   
the behavior of the quasiparticle energy spectrum.
The populations in the quasiparticle states can be controlled by using
the two-photon Bragg pulses. When the superfluid is irradiated by an external
moving Bragg potential $ V_{B}(t) $ the excited states
are populated by the quasiparticle with energy $ \hbar \omega $ and the momentum
$\hbar q$, depending on the value of $q$ and $\omega$ of the Bragg potential  
$ V_{B}(t)$. Due to the axial symmetry, the modes having only zero angular
momentum can be excited in the Bragg scattering experiments.
Since the Bragg potential is much weaker than the lattice potential, we can
treat the scattering process with linear response theory.
When the system is subjected to a time-dependent Bragg pulses, the additional
interaction term appears in the total energy functional, which is given by,
\be
H_I(t) = \int dV \psi^{\dag}(r,z,t) [V_0 \cos(qz-\omega t)]  \psi(r,z,t).
\ee
The Bragg potential can be approximated as 
$ V_B(t) = V_0 \cos( jqd-\omega t) $ since the atoms scatter off
from the condensate by the Bragg pulses from the $j$-th layer.
The above interaction energy can be re-written as
$$ \nonumber
H_I(t) = \sum_j \int d^2r \psi_j^{\dag}(r,t) V_0 \cos(jqd-\omega t) \psi_j(r,t) \nn.
$$
Following the Refs. \cite{tkg1,blak}, the momentum transfer to the Bose system placed in 
the optical lattices due the Bragg potential can be calculated analytically and 
it is given by
\bearr
P_z(t) & = & \sum_{\alpha,k} \hbar k < \hat b_{\alpha,k}^{\dag} (t) \hat b_{\alpha,k} (t) >
= \l (\frac{V_0}{2 \hbar} \r )^2  \nonumber \\
&  \times & \sum_{\alpha} \hbar q S_{\alpha} (\tilde q)
(F_{\alpha}[\omega_{-}t] - F_{\alpha}[\omega_{+}t]),
\eearr
where $ \hat b_{\alpha,k} (t) $ is the time-evolution of the quasiparticle operator
of energy $\hbar \omega_{\alpha}(k) $ and
\be
F_{\alpha}[\omega_{\pm}t] =
\l (\frac{\sin[(\omega_{\alpha}(q) \pm \omega)t/2]}{(\omega_{\alpha}(q) \pm \omega)/2} \r )^2.
\ee

In Fig. 5, we plot $ P_z(t) $ as a function of the Bragg frequency for
two values of the Bragg momenta with various time duration.
For positive $ \omega $ and a given $ \tilde q $ such that $ S_{\alpha}(\tilde q) $ is 
maximum,
the momentum transferred $ P_z(t)$ is resonant at the frequencies $ \omega = 
\omega_{\alpha}(q) $.
The width of the each peak goes like $ 2 \pi/t $. 
Figure 5 shows that shape of the $P_z(t)$ strongly depends on the time duration  
of the Bragg pulses.
When $ t = 0.5 T_r $, the $P_z(t)$ is a smooth curve for both values of the Bragg
momenta. Here, $ T_r = 2\pi /\omega_r $ is the radial trapping period.
When $t = 1.0 T_r $, there is a clear evidence of few small peaks developed in 
the $P_z(t)$.
When $ t = 2.0 T_r $, the multiple peaks in the $P_z(t)$ appears sharply.
The location of the peaks in Fig. 5 for $ t \geq 1.0 T_r $ are exactly same as 
in Fig. 4.
It implies that $P_z(t) \sim S(q,\omega)$ for a long duration of the Bragg pulses.
Therefore, in order to resolve the different
peaks, the duration of the Bragg pulses should be at least of the order of
radial trapping period $ T_{r} $.
By measuring the $ P_z $ at different Bragg momenta, it is possible to get information
about the MBBS.

\section{Summary and conclusions}
In this work, we have studied excitation energies and the corresponding
eigenfunctions of the axial quasiparticles with various discrete radial 
nodes of an array of weakly coupled quasi-two dimensional Bose condensates.
Our discretized hydrodynamic description enables us to produce
correctly all low-energy MBBS and the corresponding eigenfunctions
by including the mode couplings among different modes within the
same angular momentum sector.
We have also calculated the dynamic structure factor and the
momentum transferred to the system by the Bragg potential.
In order to resolve the multiple peaks in the dynamic structure
factor, the time duration of the Bragg pulses must be at least
of the order of the radial trapping period $ T_r$.
The MBBS can be measured by measuring $ P_z(t) $ for different
values of Bragg momenta, similar to the experiment \cite{mbs2}.
By measuring the MBBS, one may confirm the effect of the mode
coupling on the spectrum.

\begin{acknowledgments}
This work of TKG was supported by a grant (Grant No. P04311) of the
Japan Society for the Promotion of Science and partially supported
by the Alexander von Humboldt foundation, Germany.
We would also like to thank M. Ichioka for useful discussions.
\end{acknowledgments}

\end{document}